**Competing Risks Analysis on Times to Commit Crimes**

Jenq-Daw Lee
Graduate Institute of Political Economy
National Cheng Kung University
Tainan, Taiwan 70101
ROC

Cheng K. Lee
Targeting Modeling Team
Insight & Innovation
Division of Marketing
Wachovia Corporation
Charlotte, North Carolina 28244
USA

**Abstract**
A trivariate Weibull survival model using competing risks concept is applied on studying recidivism of committing 3 types of crimes – sex, violent and others. The assumption of independence of time to commit each type of crimes is relaxed so that the association of the time to recidivism between any two types of crimes can be evaluated. We found that the correlation of time to recidivism between sex crimes and violent crimes are more correlated than other pairs. Probability of experiencing a charged arrest of other crimes is greater than a charged arrest of violent crimes followed by a charged arrest of sex crimes for an individual after release.

**1. INTRODUCTION**
Survival analysis technique has been applied on criminology in finding time (also known as survival time or failure time in survival analysis) to recidivism, the probability of recidivism at a given time (also known as the survival probability in survival analysis), and the risk of recidivism at a specific time (also known as the hazard rate) given no crime activities prior to the specific time. Hill *et al*. (2008) applied Kaplan-Meier estimates, a non-parametric method, on predicting recidivism on sexual offenders. Bhati (2007) studied numbers of crimes averted by incapacitation using a semi-parametric hazard function without assuming any distribution. Loza *et al*. (2005) applied survival analysis on finding the probability of failures for 3 risk groups on reincarceration criterion. Using Cox's proportional model with covariates including some social capital factors, Liu (2005) studied time to recidivism on prisoners in China. Benda (2005) also used Cox's proportional model with demographic covariates to predict recidivism on a boot camp's graduates. Escarela *et al*. (2000) applied the competing risks concept in analyzing recidivism of three types of reconviction assuming that times to the reconviction of the three types are independent. They also had done a thorough survey on literature of applying survival analysis on criminology.

As did Escarela *et al.* (2000), in this article, these 3 types of crimes compete with each other for an individual's recidivism defined as time to a charged arrest. That is every individual after release is facing 3 events of committing sex crimes, committing violent crimes and committing other crimes. To study the times to a specific event, we let the time to commit each type of crimes be a random variable of the Weibull distribution and, therefore, a trivariate Weibull survival model is proposed. Also, some covariates are incorporated into the model to analyze the relationship with the random variables. The assumption of the independence among the 3 random variables is relaxed so that correlations between each pair of the variables can be calculated.

In Section 2, the description of the data set and the preliminary analysis are provided. The analysis using the proposed trivariate Weibull survival model is in Section 3. We conclude the article with a discussion.

## 2. DATA AND PRELIMINARY STUDY
### 2.1. The Data Set

The data studied in this article is Recidivism of Prisoners Released in 1994 downloaded from The Interuniversity Consortium for Political and Social Research (ICPSR) at the University of Michigan. The data of 38,624 observations recorded the most 3 serious charges of prisoners released in 15 states in 1994. The hierarchy from the most serious to the least serious is murder, rape, violent crimes and non-violent crimes. The ICPSR study number of this data set is 3355. After deleting observations without age at release and time served, the number of the data analyzed is 33,740.

As did Escarela *et al.* (2000), we create the same types of convicted charges for the prisoners in prison and also for the charged arrests after 1994's release. The 3 types of convicted charges and charged arrests are sex crimes, violent crimes and others. In survival analysis, the time to an event is the random variable to be studied. The event, in this article, is the recidivism defined as the first charged arrest of one of the 3 types of crimes after 1994's release. To study the association of time to the most serious charged arrests after release, we let $E_1$ be the first charged arrest of sex crimes, $E_2$ be the first charged arrest of violent crimes, $E_3$ be the first charged arrest of other crimes, $X_1$ be time to $E_1$, $X_2$ be time to $E_2$, and $X_3$ be time to $E_3$. Again, $E_1$, $E_2$, $E_3$, $X_1$, $X_2$, and $X_3$ are defined after the release date in 1994 for each individual.

As it is mentioned, the original data recorded the 3 most serious crimes of each charged arrest. We let the hierarchy of these 3 types of crimes defined by Escarela *et al.* be sex, violence and others. In our setting, $E_1$, $E_2$, and $E_3$ are the 3 events competing to each other for the occurrence to each individual after his or her release. An individual is said to be uncensored due to event $E_i$ when $E_i$ occurred, and censored when $E_i$ did not occur, $i =$ 1, 2, 3. Each individual might experience any, both, all or none of the 3 events. When $E_1$, $E_2$, or $E_3$ occurred, the corresponding $X_1$, $X_2$, and $X_3$ were observed. When $E_1$, $E_2$, or $E_3$ did not occur, $X_1$, $X_2$, and $X_3$ are set to be 1,096 days because each released individual was followed up for 3 years after the release date. There were 519 individuals arrested on the same day of release. To accommodate these observations, we set the first day of follow-up as the day of release. Table 1 shows the number of individuals with 8 combinations of the occurrences of the 3 events.

## 2.2. Preliminary Study

First, in Figure 1, we plot the cumulative hazard function using Kaplan-Meier estimates for each event assuming the 3 events are independent. The cumulative hazard curves of a constant slope suggest that $X_1$ and $X_2$ have constant hazard rates while $X_3$ has a decreasing hazard rate because of decreasing slopes in the cumulative hazard. Interpretations are that the risk of committing first charged sex crime or first charged violent crime does not increase or decrease when days passed by after release. However, the risk of committing first charged other crimec decreases when days passed by after release. Based on these findings, we choose Weibull distribution as the marginal survival function for $X_1$, $X_2$, and $X_3$ for the proposed trivariate Weibull survival function. In Weibull survival function $\exp\left[-\left(\frac{x}{\lambda}\right)^{\gamma}\right]$, the hazard rate $\frac{\gamma}{\lambda^{\gamma}} x^{\gamma-1}$ increases when the shape parameter $\gamma$ is greater than 1, the hazard rate decreases when the shape parameter $\gamma$ is less than 1, and the hazard rate remains constant when the shape parameter $\gamma$ equals 1. Figure 1 also indicates that, at any given time, individuals have the lowest risk of committing sex crimes and the highest risk of committing other crimes after release. Because the data recorded the 3 most serious charged offenses at each arrest and, based on the original data, 9,766 arrests have at least two different types offenses charged simultaneously, it is reasonable to assume that the times to commit the 3 types of crime are positively correlated.

## 3. DATA ANALYSIS USING THE TRIVARIATE WEIBULL SURVIVAL FUNCTION

Using the multivariate Weibull model by Lee and Wen (2006), the proposed trivariate Weibull survival function for $X_1$, $X_2$, and $X_3$ is constructed as

$$S_{X_1,X_2,X_3}(x_1, x_2, ..., x_n) = \exp\left\{-\left[\left(\frac{x_1}{\lambda_1}\right)^{\frac{\gamma_1}{\alpha}} + \left(\frac{x_2}{\lambda_2}\right)^{\frac{\gamma_2}{\alpha}} + \left(\frac{x_3}{\lambda_3}\right)^{\frac{\gamma_3}{\alpha}}\right]^{\alpha}\right\}$$

where $\alpha$ ($0 < \alpha \leq 1$) measures the association among the 3 random variables. $X_1$, $X_2$, and $X_3$ are independent when $\alpha$ equals 1. The scale parameters $\lambda_1$, $\lambda_2$, and $\lambda_3$ are positive as well as the shape parameters $\gamma_1$, $\gamma_2$, and $\gamma_3$. The proposed trivariate Weibull model can also be constructed using the well known Gumbel Copula with Weibull marginal. All the estimates are obtained by maximizing the log-likelihood function derived from the trivariate Weibull survival function. The likelihood function is

$$L(\boldsymbol{\theta}) = \prod_{i=1}^{N} f_{X_1,X_2,X_3}\left(t_{x_{1_i}}, t_{x_{2_i}}, t_{x_{3_i}}\right)^{p_{1_i}}$$

$$\times \left(\left[-\frac{\partial}{\partial x_1} S_{X_1,X_2,X_3}(x_1, x_2, x_3)\right]_{x_1=t_{x_{1_i}}, x_2=t_{x_{2_i}}, x_3=t_{x_{3_i}}}\right)^{p_{2_i}}$$

$$\times \left(\left[-\frac{\partial}{\partial x_2} S_{X_1,X_2,X_3}(x_1, x_2, x_3)\right]_{x_1=t_{x_{1_i}}, x_2=t_{x_{2_i}}, x_3=t_{x_{3_i}}}\right)^{p_{3_i}}$$

$$\times \left( \left[ -\frac{\partial}{\partial x_3} S_{X_1,X_2,X_3}(x_1,x_2,x_3) \right]_{x_1=t_{x_{1i}}, x_2=t_{x_{2i}}, x_3=t_{x_{3i}}} \right)^{p_{4i}}$$

$$\times \left( \left[ \frac{\partial}{\partial x_1 \partial x_2} S_{X_1,X_2,X_3}(x_1,x_2,x_3) \right]_{x_1=t_{x_{1i}}, x_2=t_{x_{2i}}, x_3=t_{x_{3i}}} \right)^{p_{5i}}$$

$$\times \left( \left[ \frac{\partial}{\partial x_1 \partial x_3} S_{X_1,X_2,X_3}(x_1,x_2,x_3) \right]_{x_1=t_{x_{1i}}, x_2=t_{x_{2i}}, x_3=t_{x_{3i}}} \right)^{p_{6i}}$$

$$\times \left( \left[ \frac{\partial}{\partial x_2 \partial x_3} S_{X_1,X_2,X_3}(x_1,x_2,x_3) \right]_{x_1=t_{x_{1i}}, x_2=t_{x_{2i}}, x_3=t_{x_{3i}}} \right)^{p_{7i}}$$

$$\times \left( \left[ S_{X_1,X_2,X_3}(x_1,x_2,...,x_n) \right]_{x_1=x_2=x_3=t_c} \right)^{1-p_{1i}-p_{2i}-p_{3i}-p_{4i}-p_{5i}-p_{6i}-p_{7i}}$$

where $f_{X_1,X_2,X_3}$ is the joint probability density function of the trivariate Weibull model, $t_{x_{1i}}$, $t_{x_{2i}}$ and $t_{x_{3i}}$ are the observed time to, respectively, $E_1$, $E_2$, and $E_3$ of $i$th individual, $t_c$ is the censoring time equal to 1,096 for all individuals with no charged arrests before the end of the study, and $p_1$, $p_2$, $p_3$, $p_3$, $p_4$, $p_5$, $p_6$, and $p_7$ are case indices corresponding to the event indices in Table 1. When $p_1$ equals 1 and all other case indices are 0, the first component of the likelihood function accounts for individuals experiencing $E_1$, $E_2$, and $E_3$. When $p_2$ equals 1 and all other case indices are 0, the second component of the likelihood function accounts for individuals experiencing only $E_1$. When $p_3$ equals 1 and all other case indices are 0, the third component of the likelihood function accounts for individuals experiencing only $E_2$. When $p_4$ equals 1 and all other case indices are 0, the fourth component of the likelihood function accounts for individuals experiencing only $E_3$. When $p_5$ equals 1 and all other case indices are 0, the fifth component of the likelihood function accounts for individuals experiencing $E_1$, and $E_2$. When $p_6$ equals 1 and all other case indices are 0, the sixth component of the likelihood function accounts for individuals experiencing $E_1$, and $E_3$. When $p_7$ equals 1 and all other case indices are 0, the seventh component of the likelihood function accounts for individuals experiencing $E_2$, and $E_3$. When all the case indices equal 0, the last component in the likelihood function accounts for individuals experiencing no events and censored at the end of the study. The values of the case indices to the corresponding cases are in Table 1. Each component is derived by extending the bivariate case of Lawless (1982).

First, the model is fitted with no covariates and the results are in Table 2. All the parameter estimates are significantly different from zero. The 3 shape parameters are less than 1 which indicates that $X_1$, $X_2$, and $X_3$ have decreasing hazard rates. That is, for a released individual after a period of time of crime free activity, the risk of committing first charged sex crime, first charged violent crime or first charged other crime decreases the next day. The association measurement $\alpha$ is 0.251 indicating some correlations among $X_1$, $X_2$, and $X_3$. Applying the formulas by Lee and Wen (2006), the correlation coefficient

between $X_1$ and $X_2$ is 0.555 (0.552, 0.559), between $X_1$ and $X_3$ is 0.529 (0.525, 0.533), and between $X_2$ and $X_3$ is 0.532 (0.528, 0.534). The 95% confidence intervals are in the parentheses using Fisher's *z* transformation formulas (Krishnamoorthy & Xia, 2007). The three pairs of correlation coefficient are between 0.4 and 0.6 which indicates that the times of committing any pair of the crimes have a moderate correlation suggested by Guldford's interpretation of correlation coefficient (Cukier & Panjwani, 2007). From the parameter estimates in Table 2, we calculate the hazard rates of $X_1$, $X_2$, and $X_3$ at *t*. The results shown in Table 3 indicate that given no charged arrests prior to *t*, for an individual released in 1994, the descending order of probability of experiencing a charged arrest is other crimes, violent crimes and sex crimes.

In order to find what variables contribute to the time of committing the first charged crime, we let each scale parameter be the exponential function of the covariates defined in Table 4. The model is then fitted with these defined covariates. Table 5 shows that all the coefficient estimates are significantly different from zero except the parameter estimate of sex treatment program and the estimate of Crime_for_1994_release_dummy2 corresponding to $X_1$. To quantify some comparisons, we use Carroll's (2003) formula to calculate hazard ratio or relative hazard for the Weibull distribution. Carroll (2003) also gave a formula to calculate the variance of hazard ratio. However, because of the different parameterizations of the Weibull function, we use the total differential technique (Fisher & Fisher, 2000) for the variance calculation. Within the same type of crime convicted before 1994's release, all the hazard ratios are calculated for each covariate while other covariates are kept constant. The results are shown in Table 6. For individuals that were jailed due to sex crimes, the interpretation of the hazard ratio of males to females is that, after the 1994's release, males have 2.552 times probability of experiencing a charged arrest of sex crimes than females on the *i*th day given that no charged arrests prior to the *i*th day. By the same interpretation, males jailed due to violent crimes have 1.863 times probability of experiencing the same crime type of charged arrest on the *i*th day given that no charged arrests prior to the *i*th day. And, males jailed due to other crimes have 1.430 times probability of experiencing the same crime type of charged arrest on the *i*th day given that no charged arrests prior to the *i*th day. For other covariates, Individuals with longer serving time have lower probability of experiencing the same crime type of charged arrest after release. Individuals with higher averaged number of arrests have higher probability of experiencing the same crime type of charged arrest after release. Non-white individuals have higher probability of experiencing the same crime type of charged arrest than white individuals after release. One interesting finding is that the probability of experiencing the same type of charged arrest after release was not lowered for individuals who participated in sex treatment programs or education programs while they were in prison.

**Conclusions**
In this article, we use a trivariate Weibull model to study the recidivism of 3 types of crimes. With the relaxation of independence assumption, we do find that there exists correlation between each pair of random variables that are times to commit the 3 types of crimes after release. By using the trivariate model, not only are we able to calculate the hazard rates of the 3 random variables that a released individual was facing at any given day, but also we are able to compare the hazard rates of a released individual within the same type of crimes with different covariates. The trivariate model can be extended to a multivariate model with a higher dimension when more types of crimes are defined.

U.S. Department of Justice, Bureau of Justice Statistics. (2002). *Recidivism of Prisoners Released in 1994*, Washington, DC.

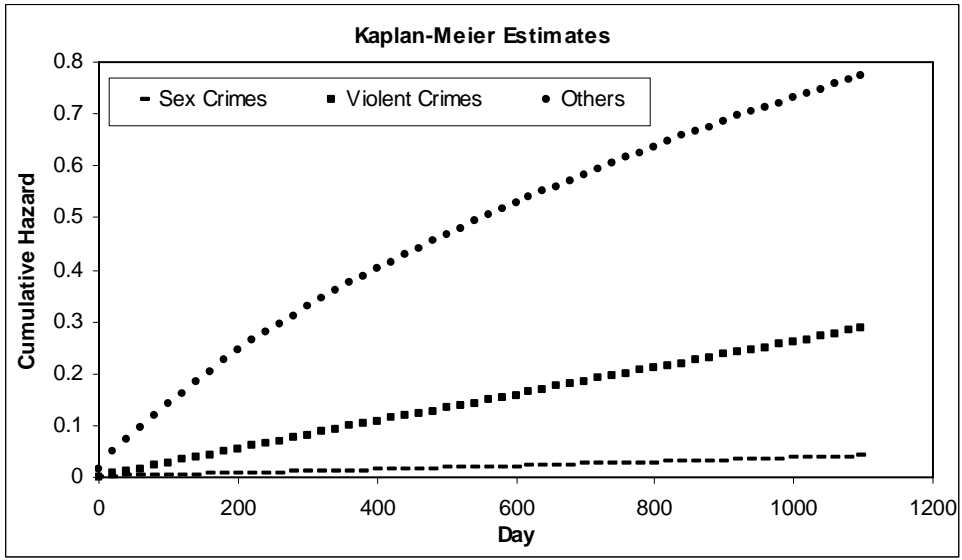

Fig 1. Kaplan-Meier Cumulative Hazard of $X_1$, $X_2$, and $X_3$.

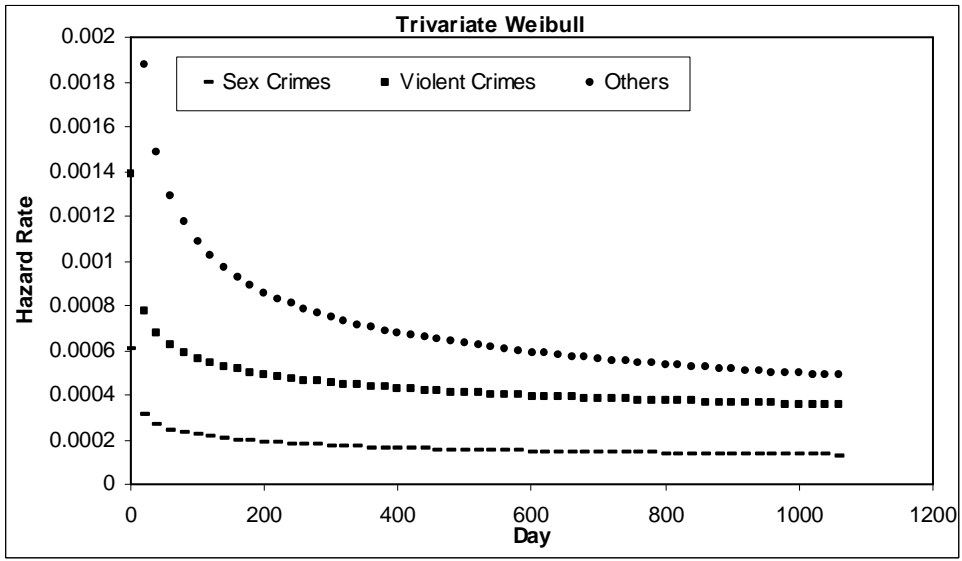

Fig 2. Trivariate Weibull Hazard of $X_1$, $X_2$, and $X_3$.

Table 1

| Case | Number | Event Occurred | Case Indices |
|---|---|---|---|
| 1 | 114 | $E_1, E_2, E_3$ | $p_1=1; p_2=0; p_3=0; p_4=0; p_5=0; p_6=0; p_7=0;$ |
| 2 | 402 | $E_1$ | $p_1=0; p_2=1; p_3=0; p_4=0; p_5=0; p_6=0; p_7=0;$ |
| 3 | 2006 | $E_2$ | $p_1=0; p_2=0; p_3=1; p_4=0; p_5=0; p_6=0; p_7=0;$ |
| 4 | 13177 | $E_3$ | $p_1=0; p_2=0; p_3=0; p_4=1; p_5=0; p_6=0; p_7=0;$ |
| 5 | 62 | $E_1, E_2$ | $p_1=0; p_2=0; p_3=0; p_4=0; p_5=1; p_6=0; p_7=0;$ |
| 6 | 287 | $E_1, E_3$ | $p_1=0; p_2=0; p_3=0; p_4=0; p_5=0; p_6=1; p_7=0;$ |
| 7 | 3852 | $E_2, E_3$ | $p_1=0; p_2=0; p_3=0; p_4=0; p_5=0; p_6=0; p_7=1;$ |
| 8 | 13840 | None | $p_1=0; p_2=0; p_3=0; p_4=0; p_5=0; p_6=0; p_7=0;$ |

Table 2

| Parameter | Estimate | 95% Confidence Interval |
|---|---|---|
| $\alpha$ | 0.475 | (0.465, 0.484) |
| $\gamma_1$ | 0.777 | ( 0.747, 0.808) |
| $\lambda_1$ | 9926.352 | (8722.389, 11130.315) |
| $\gamma_2$ | 0.804 | (0.789, 0.819) |
| $\lambda_2$ | 2742.914 | (2652.181, 2833.647) |
| $\gamma_3$ | 0.659 | (0.651, 0.668) |
| $\lambda_3$ | 1543.809 | (1507.125, 1580.493) |

Table 3

| Random Variable | Hazard Rate | Variance |
|---|---|---|
| $X_1$ | $\dfrac{0.000608}{t^{0.223}}$ | $\dfrac{2.694\times10^{-9} - 9.472\times10^{-10}\times\log(t) + 9.032\times10^{-11}\times\log^2(t)}{t^{0.445}}$ |
| $X_2$ | $\dfrac{0.00138}{t^{0.196}}$ | $\dfrac{3.614\times10^{-9} - 1.207\times10^{-9}\times\log(t) + 1.076\times10^{-10}\times\log^2(t)}{t^{0.392}}$ |
| $X_3$ | $\dfrac{0.00521}{t^{0.341}}$ | $\dfrac{1.566\times10^{-8} - 5.508\times10^{-9}\times\log(t) + 5.352\times10^{-10}\times\log^2(t)}{t^{0.681}}$ |

Table 4

| parameter | variable | Description |
|---|---|---|
| $\lambda_1$ | sex_dummy | male=1; female=0 |
| | log_tmsrv | logarithm of time served in month for 1994 imprisonment |
| | average number of arrests by age | average number of arrests divided by age at release |
| | race | white=0; non white=1; |
| | sextrt | participated in sex treatment program=1; not participated in sex treatment program=0; |
| | crime_for_1994_release_dummy1 crime_for_1994_release_dummy2 | sex crimes: crime_for_1994_release_dummy1=1 and crime_for_1994_release_dummy2=0 violent crimes: crime_for_1994_release_dummy1=0 and crime_for_1994_release_dummy2=1 other crimes: crime_for_1994_release_dummy1=0 and crime_for_1994_release_dummy2=0 |
| $\lambda_2$ | sex_dummy | male=1; female=0 |
| | log_tmsrv | logarithm of time served in month for 1994 imprisonment |
| | average number of arrests by age | average number of arrests divided by age at release |
| | race | white=0; non white=1; |
| | educat | participated in education program=1; not participated in education program=0; |
| | crime_for_1994_release_dummy1 crime_for_1994_release_dummy2 | sex crimes: crime_for_1994_release_dummy1=1 and crime_for_1994_release_dummy2=0 violent crimes: crime_for_1994_release_dummy1=0 and crime_for_1994_release_dummy2=1 other crimes: crime_for_1994_release_dummy1=0 and crime_for_1994_release_dummy2=0 |
| $\lambda_3$ | sex_dummy | male=1; female=0 |
| | log_tmsrv | logarithm of time served in month for 1994 imprisonment |
| | average number of arrests by age | average number of arrests divided by age at release |
| | race | white=0; non white=1; |
| | educat | participated in education program=1; not participated in education program=0; |
| | crime_for_1994_release_dummy1 crime_for_1994_release_dummy2 | sex crimes: crime_for_1994_release_dummy1=1 and crime_for_1994_release_dummy2=0 violent crimes: crime_for_1994_release_dummy1=0 and crime_for_1994_release_dummy2=1 other crimes: crime_for_1994_release_dummy1=0 and crime_for_1994_release_dummy2=0 |

Table 5.

| Parameter | | Estimate | 95% Confidence Interval |
|---|---|---|---|
| $\alpha$ | | 0.522 | (0.512, 0.533) |
| $\gamma_1$ | | 0.812 | (0.780, 0.844) |
| $\lambda_1$ | Intercept | 10.994 | (10.508, 11.480) |
| | Sex | -1.154 | (-1.587, -0.720) |
| | Logarithm of_tmsrv | 0.069 | (0.024, 0.115) |
| | Average number of arrests by age | -1.597 | (-1.817, -1.377) |
| | Race | -0.184 | (-0.280, -0.089) |
| | Sex treatment program | -0.073 | (-0.496, 0.350) |
| | Crime_for_1994_release_dummy1 | -0.600 | (-0.718, -0.481) |
| | Crime_for_1994_release_dummy2 | -0.134 | (-0.288, 0.021) |
| $\gamma_2$ | | 0.871 | (0.855, 0.887) |
| $\lambda_2$ | Intercept | 8.821 | (8.693, 8.948) |
| | Sex | -0.714 | (-0.820, -0.608) |
| | Logarithm of_tmsrv | 0.138 | (0.116, 0.159) |
| | Average number of arrests by age | -1.834 | (-1.909, -1.759) |
| | Race | -0.443 | (-0.487, -0.399) |
| | Education program | -0.145 | (-0.198, -0.092) |
| | Crime_for_1994_release_dummy1 | 0.397 | (0.338, 0.455) |
| | Crime_for_1994_release_dummy2 | -0.205 | (-0.260, -0.150) |
| $\gamma_3$ | | 0.706 | (0.697, 0.715) |
| $\lambda_3$ | Intercept | 7.858 | (7.754, 7.962) |
| | Sex | -0.507 | (-0.594, -0.420) |
| | Logarithm of_tmsrv | 0.178 | (0.157, 0.198) |
| | Average number of arrests by age | -2.152 | (-2.220, -2.084) |
| | Race | -0.396 | (-0.437, -0.354) |
| | Education program | -0.243 | (-0.292, -0.193) |
| | Crime_for_1994_release_dummy1 | 0.656 | (0.600, 0.711) |
| | Crime_for_1994_release_dummy2 | 0.113 | (0.058, 0.167) |

Table 6

| Jailed and rearrested after release due to sex crime | Hazard Ratio | 95% Upper Limit | 95% Lower Limit |
|---|---|---|---|
| Males vs. females | 2.552 | 2.415 | 2.690 |
| Log of Tmsrv increased by 1 unit | 0.945 | 0.906 | 0.984 |
| Number of arrests over age at release increased by 1 unit | 3.658 | 3.610 | 3.706 |
| Non-white vs. White | 1.161 | 1.095 | 1.228 |
| Participated sex treatment program vs. did not | 1.061 | 0.737 | 1.384 |

| Jailed and rearrested after release due to violent crime | Hazard Ratio | 95% Upper Limit | 95% Lower Limit |
|---|---|---|---|
| Males vs. females | 1.863 | 1.813 | 1.912 |
| Log of Tmsrv increased by 1 unit | 0.887 | 0.866 | 0.908 |
| Number of arrests over age at release increased by 1 unit | 4.939 | 4.927 | 4.952 |
| Non-white vs. White | 1.471 | 1.445 | 1.497 |
| Participated education program vs. did not | 1.134 | 1.094 | 1.175 |

| Jailed and rearrested after release due to other crime | Hazard Ratio | 95% Upper Limit | 95% Lower Limit |
|---|---|---|---|
| Males vs. females | 1.430 | 1.387 | 1.473 |
| Log of Tmsrv increased by 1 unit | 0.882 | 0.866 | 0.899 |
| Number of arrests over age at release increased by 1 unit | 4.571 | 4.561 | 4.581 |
| Non-white vs. White | 1.322 | 1.300 | 1.344 |
| Participated education program vs. did not | 1.187 | 1.158 | 1.216 |